# *What are the attitudes and beliefs about Science of the Physics teachers and future Physics teachers in Uruguay?*

## Álvaro Suárez, Daniel Baccino,

## Martín Monteiro, Arturo C. Martí

Álvaro Suárez

ORCID https://orcid.org/0000-0002-5345-5565

Consejo de Formación en Educación, Instituto de Profesores Artigas, Montevideo, Uruguay

**alsua@outlook.com**

Daniel Baccino

ORCID https://orcid.org/0000-0001-5572-2623

Consejo de Formación en Educación, Instituto de Profesores Artigas, Montevideo, Uruguay

**dbaccisi@gmail.com**

Martín Monteiro

https://orcid.org/0000-0001-9472-2116

Universidad ORT Uruguay, Montevideo, Uruguay

**monteiro@ort.edu.uy**

Arturo C. Martí

https://orcid.org/0000-0003-2023-8676

Instituto de Física, Facultad de Ciencias, Udelar, Montevideo, Uruguay

**marti@fisica.edu.uy**

**Abstract:** We investigate the epistemological conceptions of the uruguayan Physics teachers and future Physics teachers through the application of the CLASS test (Colorado Learning Attitudes about Science Survey), one of the most accepted instruments in the research community in Physics education. The results obtained allow us to compare the attitudes and beliefs about science of both groups and quantitatively evaluate the agreement or not with the conceptions of experts in the field. First, we present a general screenshot of the responses and then we identify categories in which there are significant similarities or differences between the two groups studied and in turn with the reference of the experts. The categories that show significant positive or negative variations between the opinions of future teachers and teachers indicate the areas where training is favorable or unfavorable. On the other hand, the areas where the differences with the opinions of the experts are globally notorious suggest that they should be strengthened. To get a more global perspective, we also compare our results with those published in the literature. Finally, we raise several questions that we think may favor further inquiries.

1. **Introduction.**

In 1964, the renowned American psychologist B. F. Skinner noted that "education is what survives when a man has forgotten all he has been taught". This phrase illustrates that the specific contents we teach are only a small part of the transformation that takes place as students make their way through the educational system. Indeed, students are not "empty boxes" in which teachers dump knowledge. On the contrary, each student comes with a whole baggage of knowledge of the subject, his or her own personal history, a particular way of interacting with peers and teachers, an approach to learning and a set of varied expectations, among other things. These aspects —often underestimated in relation to the curricular ones— play an important role in the transformation we expect students to undergo.

The set of ideas, assumptions and previous conceptions about Science, especially its evolution, its methods, its validation or refutation, are encompassed in the concept of *epistemological beliefs*. These beliefs, which are generally not explicit, have a strong impact on the way we teach and learn (Redish, Saul, & Steinberg, 1998). To mention a few aspects, do we consider science as a body of knowledge set in stone or as a system in constant evolution? Is the role of experimentation simply a requirement to verify theories or rather a substantial part of understanding nature, refuting theories or delimiting the range of their validity? There is no neutral stance towards these questions. If we consider previous knowledge as immovable, there is no need to foster a critical attitude or take an interest in new developments. Likewise, if we think that "anything goes" or that experimental evidence is not relevant, we fall into the field of pseudoscientific beliefs, or positions such as those held by terraplanists or antivaccinationists.

As teachers, we are directly or indirectly influencing the attitudes and beliefs of our students in the learning and teaching of Physics with every action or omission we make in class. Let us consider, for example, a hypothetical situation in a class on kinematics where we solve a problem on the whiteboard. Imagine that, as part of the solving procedure, we write down the equations describing an MRUV and tell students that, in order to find the answer, they must "find the equation that has all the data except the requested variable and simply solve it". In a hypothetical situation such as the one described, we would be unintentionally giving the student an erroneous image of the way in which physics problems should be solved. In that sense, we are (unwillingly) projecting an incorrect image of science and how to learn physics.

The aspects that frame the specific scientific knowledge of physics within society deserve to be discussed, analyzed and made explicit in science classes. While it is true that philosophy as a subject is present in most curricula, several aspects of its relationship with physics should be discussed alongside its more specific contents. Is it possible to talk about electromagnetic induction without mentioning the role of motors and generators in our daily life? Can we present quantum mechanics as an abstract theory while holding in our hands a smartphone with millions of transistors that are governed precisely by the laws of quantum mechanics? Is it reasonable to study the details of uranium fission without mentioning the social and ethical implications of the atomic bomb? These examples show how closely the contents we study and the set of beliefs and attitudes towards science are linked.

There are other more subtle, less conspicuous aspects of the attitudes toward science that are discussed even less in our classes. Some of these involve attitudes towards minorities or genders, especially towards the role of women in relation to the "hard sciences". Why is the percentage of women choosing these disciplines lower? Is the teaching we provide "neutral" towards these aspects? At what point in training does this apparent lack of interest originate? Does it occur in the early years or in more advanced stages? Is it due to the prevalent working conditions? Is it influenced by way scientists as typically portrayed in our society? Is it a matter of a "No Girls Allowed" style discrimination? All of these questions are becoming increasingly more important today. In order to move towards a convincing answer, it is only natural to delve into the epistemological attitudes and beliefs of students and teachers.

Attitudes and beliefs have multiple dimensions. One of them is transversal and weaves across the different collectives, from the attitudes of society, science students, future physics teachers and current ones to those of researchers working in this discipline. In this paper, we focus on evaluating and discussing the attitudes and beliefs of high school physics teachers and physics-teacher-training students in Uruguay. In the next section, we present a set of tools developed in recent years to analyze these aspects, which constitute our theoretical framework, and we provide the most relevant results obtained from their application in different contexts. Regarding the attitudes and beliefs of Uruguayan teachers at the core of this study, in section we present the methodology used and in section 4, the main results. Finally, the discussion and final considerations are presented in section 5.

## 2. Tools for the analysis of epistemological beliefs and their main results.

### 2.1 Standardized Questionnaires in Physics Education Research

Since nearly half a century ago, in has been observed that students in different fields of physics frequently learn to solve problems and pass their courses, but, when faced with simple conceptual problems, they often lack a thorough understanding of the studied phenomena (Viennot, 1979; Trowbridge, & McDermott, 1980; Halloun, & Hestenes, 1985; Hake, 1996, Docktor, & Mestre, 2014). Based on these observations, a demand for changes rose from the core of physics. Studies on this topic consolidated over time and a new field with its own identity emerged: Physics Education Research (PER) (McDermott, 1999), which aims to generate knowledge based on widely accepted scientific methods and especially on quantitative results with solid statistical support (Docktor, & Mestre, 2014).

One of the pillars of PER is the systematic evaluation of knowledge. Within this framework, several questionnaires were devised for the evaluation of knowledge, including the Force Concept Inventory (Hestenes, Wells, & Swackhamer, 1992). This questionnaire, applied to thousands of students in hundreds of universities, aims to evaluate the transition from an Aristotelian to a Newtonian conception of mechanics. The main conclusion derived from the analysis of the results is that active teaching methods, in which students are actively involved, result in better learning compared to traditional methods, mainly consisting of lectures where students have a passive attitude. This greater "gain" in learning is a statistically robust result that does not depend on the student, the teacher or the context in which the learning occurs (Hake, 1998).

In the years that followed, several other questionnaires were proposed, such as the *Conceptual Survey of Electricity and Magnetism* (CSEM) (Malonney et al., 2001) or the *Brief Electricity and Magnetism Assessment* (BEMA) (Chabay, & Sherwood, 2006), which focus on other fields of physics. While in the early stages these questionnaires were relatively handcrafted, over time a series of steps, statistical requirements and stages that each proposal should fulfill to be accepted in the community were defined (Heron, & Meltzer, 2005). Several collaborative sites (notably http://www.physport.org) publish the questionnaires, generally with open access, and complementary information including links to the articles such, translations into many languages, spreadsheets and other tools to facilitate their implementation. On the other hand, when those requirements are met, it is possible to share information, both by uploading our own results and by accessing those of other institutions.

### 2.2 Analysis of epistemological attitudes and beliefs.

The field of epistemological attitudes and beliefs does not escape the researchers' aspiration to obtain quantitative information. In that sense, a series of questionnaires or standardized tests have been developed in the last twenty-five years to assess students' attitudes and epistemological beliefs about physics, its teaching and learning. Unlike specific knowledge tests, which are mostly based on multiple choice questions, attitudes and beliefs tests usually ask students to indicate the degree to which they agree or disagree (known as a Likert scale) with statements that reflect (or not) the opinion of "experts" in the discipline (i.e., professional physicists, researchers). Each statement has an expected value assigned during the design and validation process of the instrument, based on repeated interactions with experts. The differences between the students' answers and the experts' answers constitute the raw material for analyzing the epistemological status of the group of students. In general, it is of interest to analyze the impact that particular courses and teaching approaches have on students' attitudes and beliefs. To this end, the test is applied at least twice, the first time at the beginning of the course (pretest) and the second time at the end of the course (posttest). The difference between the two tests allows us to measure changes in the attitudes and beliefs as a function of demographics, methodologies and teaching strategies.

Within the set of standardized tests, three of them are of particular relevance: MPEX (Maryland Physics Expectations Survey) (Redish, et al., 1998), which, as its name indicates, aims to survey students' expectations in relation to physics, CLASS (Colorado Learning Attitudes about Science Survey) (Adams, et al., 2006), which aims to assess students' beliefs about physics and their learning, and E-CLASS (Colorado Learning Attitudes about Science Survey for Experimental Physics) (Wilcox, & Lewandowski, 2016), with the same guiding principles as CLASS but focused on the experimental aspects of physics.

## 2.3. Teaching methods and epistemological attitudes

Numerous papers have been published about PER, exploring the relationship between teaching methods and outcomes and the epistemological attitudes of students (Madsen, McKagan, & Sayre, 2015). In particular, it has been shown that certain "negative" epistemological attitudes of students are definitely an obstacle to the development of quality science education, as they affect the way students learn and approach science courses (Perkins, et al., 2005; Milner-Bolotin, Antimirova, Noack, & Petrov, 2011). For example, the way in which a student believes he/she should learn physics, his/her metacognitive practices, as well as the image he/she has of physics as a science, are some of the aspects of the hidden curriculum that should affect the way he/she learns. A student who believes that physics consists mainly of disconnected facts and formulas will study differently from one who sees it as a network of interconnected concepts. Those who view physics knowledge as a coherent network of ideas have more reason to engage in metacognitive practices to monitor their learning (Redish, Saul, & Steinberg, 1998). Regarding academic performance, several studies have shown the existence of a positive correlation with CLASS and MPEX scores (Perkins, et al., 2005; Cahill, et al., 2018). In recent years, there have been multiple quantitative investigations on the changes that occur in students' attitudes and beliefs as a result of their previous training, types of courses and teaching strategies used, as well as the relationship between pretest and academic performance, among other variables (Madsen, McKagan, & Sayre, 2015).

Among all the research done on the subject, we would particularly like to highlight the studies that focus on the changes in epistemological attitudes as a result of teaching methods, as well as those that analyze the relationship between these attitudes and conceptual understanding. Based on a meta-analysis consisting of the study of 24 published research papers, Madsen, McKagan and Sayre (2015) classified teaching strategies in physics courses into three categories based on changes in the epistemological attitudes of students:

A) Courses where there is a decline in students' attitudes and beliefs (the average posttest score on attitudes and beliefs is lower than the average pretest score). This category includes courses with traditional teaching methodologies, as well as those developed with some of the more widespread active-teaching methodologies, such as "Peer Instruction" and "Introductory Physics Tutorials" (Adams et al, 2006; Madsen, et al., 2015). While the notion of traditional courses not promoting the right attitudes and beliefs could be reasonably surmised, it is striking at first glance that students also underperform on CLASS and MPEX after taking courses based on active teaching strategies.

B) Courses where no significant difference is detected in students' attitudes and beliefs before and after the course. This group consists of courses where some epistemological aspects are taken into account in their implementation (Kohl, & Vincent Kuo, 2012; Madsen, et al., 2015). These courses implement teaching strategies that promote the development of reasoning skills, inference-making from observation, and reflections on why we trust scientific ideas, among others.

C) Courses where a significant improvement in students' attitudes and beliefs is detected. There are two types of courses in this group: those that are radically restructured, based on model building (Brewe, et al., 2013), and those with a strong emphasis on epistemological aspects (Elby, 2001; Redish, & Hammer, 2009). In courses based on model building, students work in small groups, perform experiments and analyze the results in order to develop models of different phenomena. A central aspect of this type of strategy is the sharing of each team's conclusions and the promotion of peer discussion to allow students to validate and refine the models they build. On the other hand, courses with a defined epistemological approach develop activities that promote metacognition, for example, working with tutorials that emphasize the integration of intuitive ideas and formal scientific thought.

Therefore, as supported by the literature, the attitudes and beliefs of students generally decay as a collateral effect of physics courses, except in courses specifically designed to promote an improvement in those attitudes and beliefs. Surprisingly, at the end of many physics courses, more students believe, for example, that physics is less connected to the real world and less coherent, that reasoning is less important, and that rote learning is useful. These results, as shocking as they may be, pose the question about the unintentional promotion of attitudes and beliefs contrary to those that we would expect to develop naturally by simply taking the courses. The close relationship between epistemological attitudes and the results of the educational process justifies to a great extent the need to understand and act on the attitudes and beliefs of students throughout the whole educational system.

## 3. Methodology

In the research presented in this paper we chose to use CLASS, the science learning attitudes survey developed by the PER group at the University of Colorado Boulder, in the United States (Adams, et al., 2006). This questionnaire is widely accepted as it has undergone an exhaustive validation process. It was developed based on tools such as MPEX, and then reviewed based on interviews with students and experts (16 physicists with extensive teaching experience who agreed with most of the answers). CLASS categories were created by means of a reduced-basis factor analysis in which raw statistical categories and researcher-determined categories were iteratively combined. Finally, it was applied to thousands of students and we verified that the beliefs of those with more experience in physics closer to those of the experts. The results of their application in different geographical and social contexts, or in groups where different teaching strategies are applied, are available for analysis and comparison (Milner-Bolotin, Antimirova, Noack, & Petrov, 2011; Ding, 2013; Suwonjandee, Mahachok, & Asavapibhop, 2018; Balta, Cessna, & Kaliyeva, 2020, Kontro, & Buschhüter, 2020; Nissen et al., 2021). This instrument assesses students' attitudes towards learning physics, how they think physics relates to their daily lives, and what they think about physics. It is comprised of 42 statements, each with a five-level Likert scale, from Completely disagree (1) to Completely agree (5). We note some questions as examples:

- *"I think about the physics involved in my everyday life."*
- *"Knowledge in physics consists of many disconnected topics."*
- *"After I study a physics topic and feel I understand it, I have difficulty solving problems on the same topic."*
- *"When solving a physics problem, I look for an equation that uses the variables given in the problem and I substitute the values."*

The statements can be organized, according to the authors of CLASS, into eight non-exclusive categories, as shown in Table 1. When processing a survey taken by a student, each of his or her answers will receive a value (-1 or +1), depending on the degree of disagreement (-1) or agreement (+1) between the answer given by the student and the typical answer of the experts. This criterion only applies to the 36 questions on which there is real consensus among experts (only 27 of the 36 are categorized). On the contrary, it does not apply to the six questions on which experts do not agree (see the last row of Table 1). Based on these values, the levels of agreement ("favorable" and "unfavorable") are determined for each of the eight categories defined by the authors, as well as for the survey as a whole. After applying this process to all the individuals in the surveyed group, we calculated the averages for each statement, category and survey total.

| Category | Statements | ID |
|---|---|---|
| Real world connection | 28, 30, 35, 37 | RWC |
| Personal interest | 3, 11, 14, 25, 28, 30 | PI |
| Sense making / Effort | 11, 23, 24, 32, 36, 39 | SM/E |
| Conceptual comprehension | 1, 5, 6, 13, 21, 32 | CC |
| Applied conceptual understanding | 1, 5, 6, 8, 21, 22, 40 | ACU |
| Problem solving, General | 13, 15, 16, 25, 26, 34 | PSG |
| Problem solving, Confidence | 15, 16, 34, 40 | PSC |
| Problem solving, Sophistication | 5, 21, 22, 25, 34, 40 | PSS |
| No agreement among experts | 4, 7, 9, 31, 33, 41 | |

**Table 1. Identification of categories and the statements included in each category (Adams et al, 2006, translated by the authors).**

The test was implemented electronically through a form sent to active physics teachers working in secondary education and first year students of the physics teacher-training program of the CFE (Uruguayan Council for Training in Education). The survey was open to every collective, teachers and students throughout Uruguay for one month during the first half of 2020.

We received 143 responses from teachers. The gender stated by the respondents was as follows: Female (46% of the total), Male (52%), 2% did not state their gender. Fifty-two percent of the respondents stated that they carry out their teaching activities in Montevideo (the country's capital), while the remaining 48% are distributed in other departments of the country. As for the age of the respondents, 31% stated to be between 31 and 40 years old, 22% stated to be in the subsequent age ranges (18-30 and 41-50), while 25% stated to be over 51[1].

A total of 138 students from almost all the centers where physics is taught answered the survey. As for gender, 62% of respondents stated that they were female, 38% stated that they were male, and no respondents chose the other available options. In terms of the institute where they are based, most of them (56%) study in a blended-learning model[2], while the rest study on-site at different centers in the country.

The results of each of the surveys were analyzed based on the categories defined and validated by the authors of the instrument. With this as our starting point, we used an electronic spreadsheet taken from the Physport portal and adapted by the authors of this work to quantify the results. The following section shows the main results.

## 4. Analysis of results

4.1 Teachers and teacher-training students in Uruguay

In a first analysis of the results, we examine all the categorized responses of teacher. The histogram in Figure 1 represents the frequencies (number of teachers) whose answers are in agreement with the experts, in intervals of 10 percentage points wide. For example, a total of 56 teachers gave "favorable" responses that are between 80% and 90%. The typical way to characterize this distribution is by the average, which in this case is 80%, the standard deviation (13%) and the standard deviation of the average (1%).

---

1 The results of the teacher and student survey are available at bit.ly/CLASS-profes-UY2020 and bit.ly/CLASS-formacion-UY2020.
2 Blended learning consists of a model of teacher training available since 2003 at Teacher Training Centers throughout Uruguay, for which specific subjects are not available in some specialties, one of them being physics.

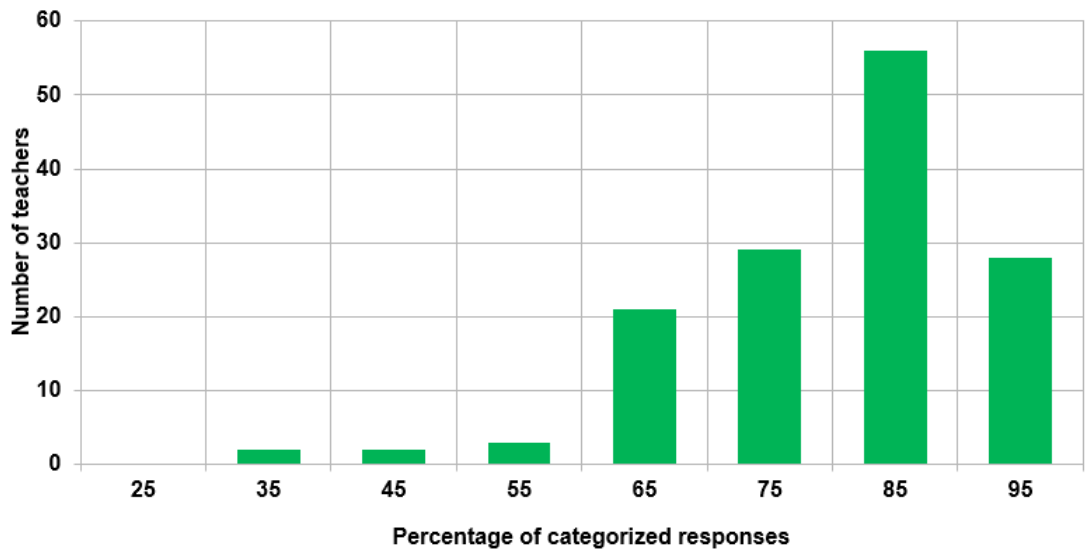

Figure 1. Overall results of the responses to the questions categorized in the CLASS test by 143 physics teachers in Uruguay.

In Figure 2 we show the results by categories in the teacher survey, for favorable, neutral and unfavorable responses (according to the ID in Table 1); the first two columns represent the percentages corresponding to all responses (All) and total categorized responses (All C.), respectively.

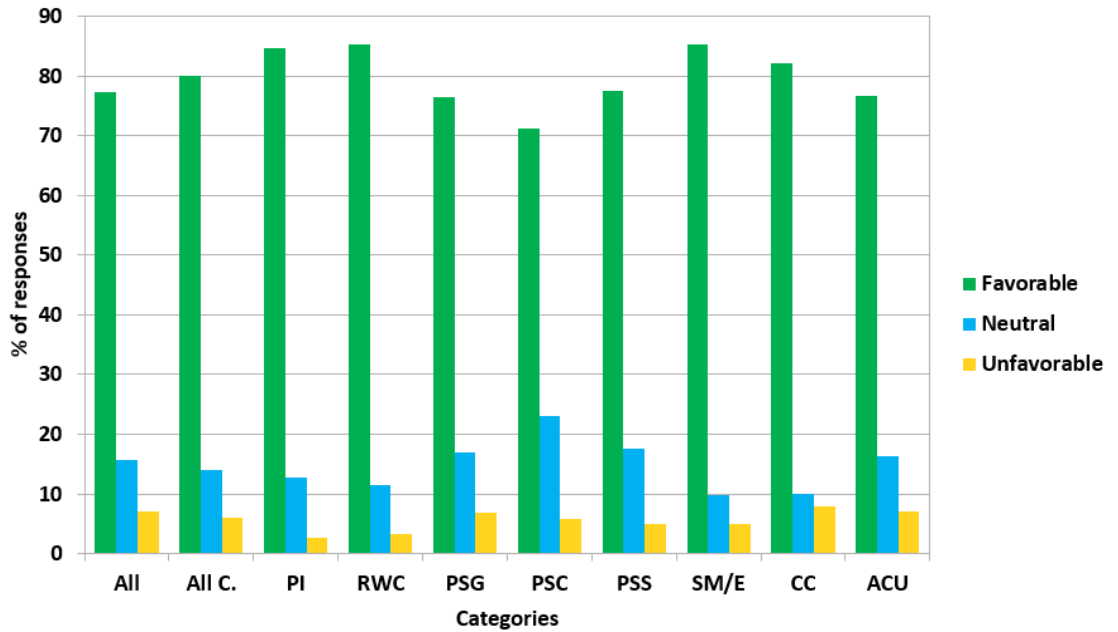

Figure 2. Performance of physics teachers surveyed with the CLASS test broken down by category.

The graph in Figure 3 shows the results of favorable responses of physics teachers and teacher-training students broken down by CLASS categories (the same comments regarding the identification of categories made on the graph in Figure 2 apply here). For each category and in both groups, the standard error of the average is shown. The percentage axis is defined between 45% and 95%, in order to better visualize the area of interest.

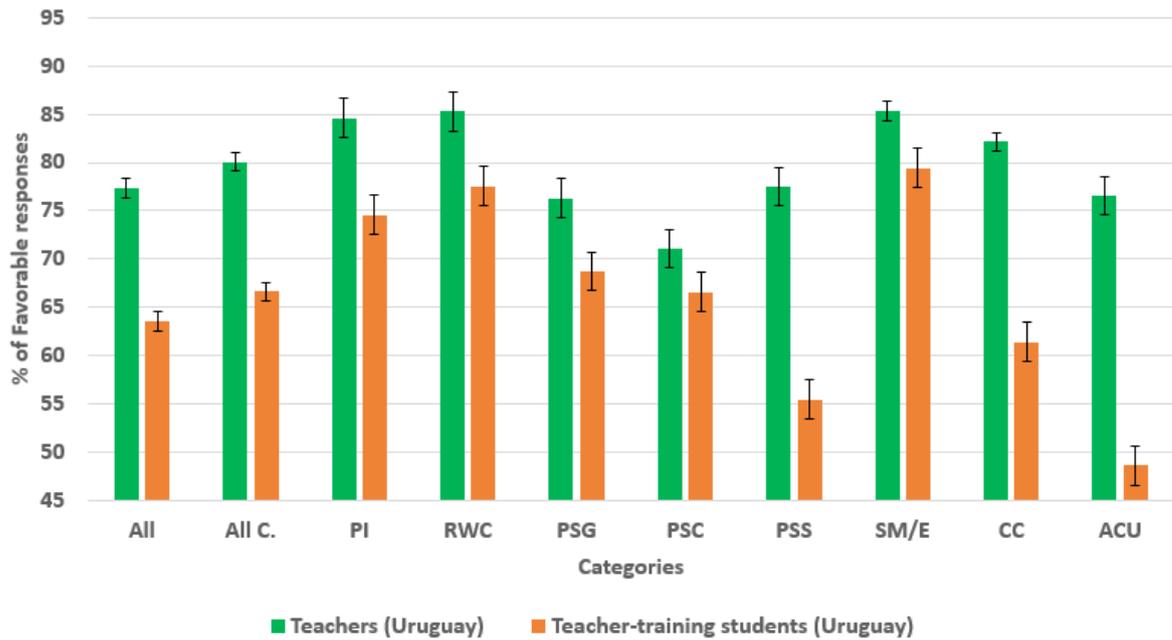

Figure 3. Performance of 138 physics teacher-training students and 143 active physics teachers on the CLASS test broken down by category.

One aspect that stands out in the graph is the difference in the averages of teacher and student responses. The average of categorized responses is 80% for teachers and 67% for students. If we consider the range defined by the averages of all categories, we observe that it is approximately 30 percentage points in the case of students, whereas in the case of teachers it is approximately 15 points. In the case of teachers, the responses to the categories are less dispersed than in the subsets categorized for students.

The categories that show the most agreement with the experts are the same for students and teachers. In some categories, the agreement of students and teachers with experts is very similar, for example, in *Sense making/ Effort* and *Confidence in problem solving*. On the other hand, the greatest differences are observed in the categories *Applied conceptual understanding* and *Sophistication in problem solving*. Assuming that it is possible to consider the two groups of surveyed individuals as part of a pseudo-longitudinal approach, these differences could be hypothetically explained by the systematic training processes and those mediated by the teachers' daily practice.

We proposed the survey to 79 students pursuing a Bachelor's Degrees in Physics and Mathematics at the Faculty of Sciences (University of the Republic) and to 105 students enrolled at the Faculty of Engineering of the same university. We found no significant differences between the responses of the undergraduate students and the physics teacher-training students surveyed. The same survey was presented to local researchers in the area of Physical Science. Again, we found no significant differences when comparing the results of the latter with those recorded among physics teachers. The analysis results of the responses of these groups will be analyzed as part of a subsequent study.

### 4.2 Comparison with international results

It is interesting to compare the overall results of physics teacher-training students in Uruguay with university students from other parts of the world. Using data from the meta-analysis by Madsen, McKagan and Sayre (2015) and Bates, Galloway, Loptson and Slaughter (2011), we took the average values of the implementation of the CLASS test among students from thirteen universities in Canada, the United States and the United Kingdom, considering the fact that those students take calculus-based physics courses as part of their curricula. This point is significant, since students who pursue majors with a strong emphasis on physics and mathematics perform better on CLASS according to the literature (Madsen, McKagan, & Sayre, 2015).

While the average of the CLASS test among teacher-training students for the 36 questions where there is agreement among experts was 63.5%, the average for the thirteen surveyed groups is 64.1% with a standard deviation of 6.0%. These results show that physics teacher-training students in Uruguay obtain results similar to those of the populations used for comparison. Although they do not explain anything about education systems, these results suggest that there may be a trend in the average CLASS test scores for students with science and engineering profiles, regardless of their country of origin.

On the other hand, regarding the attitudes and beliefs of physics teachers, a comparison with a study conducted in Thailand, a country with very different demographic, social, economic and cultural characteristics than Uruguay, may be constructive. Said study, conducted by Suwonjandee, Mahachok, and Asavapibhop (2018), analyzes the responses of 196 physics teachers and 211 high school students. Despite the aforementioned contextual differences, the responses were very similar in most CLASS questions and categories. Only two of the 42 test questions show significant differences between teachers from Thailand and Uruguay (questions 8 and 22). These are two of the six questions in the *Applied conceptual understanding* category.

In the aforementioned questions, teachers in Uruguay showed a higher degree of agreement with the experts than teachers in Thailand, something very similar to what happens with physics teacher-training students in Uruguay, who also show great differences with the experts precisely in those same two questions. In fact, it may be relevant to note that these questions are the ones in our research that present the highest degree of discrepancy between the responses of physics teachers and teacher-training students:

- Question 8: *"When solving a physics problem, I look for an equation that uses the variables given in the problem and I substitute the values."* The degree of agreement with the experts was 60% and 24% for Uruguayan teachers and teacher-training students, respectively, and 9% for teachers from Thailand.
- Question 22: *"If I want to apply a method used in one physics problem to a different one, these problems must involve very similar situations."* The degree of agreement with the experts was 67% and 27% for Uruguayan teachers and teacher-training students, respectively, and 38% for teachers from Thailand.

These two questions seem to point out that the attitude towards physics problems is a key aspect in the training of physics teachers and seem to be a relative strength of Uruguayan teachers. The low degree of agreement shown by Uruguayan teacher-training students in question 8 seems to suggest that they approach physics problems by plugging in equations, solving them with the available data, without thinking too much about the conceptual aspects. Question 22 seems to reinforce this idea, implying that teacher-training students fail to fully grasp the conceptual aspects and basic principles of physics and instead learn physics by memorizing equations, which would prevent them from approaching new situations with enough confidence to solve them.

## 5. Final comments

The epistemological attitudes and beliefs of teachers and students play a fundamental role in the classroom, involving many different aspects. In that sense, it is essential to be aware of its importance and recognize the impact that our actions within the classroom can have on the image of science that we promote, the choice of future careers or the academic achievements of our students.

Students' "epistemological" beliefs (their views on the nature of knowledge and learning) are an obstacle to the development of quality science education, affecting the way students learn and approach science courses. For example, those who consider that learning consists mainly of absorbing information will have a different attitude towards knowledge than those who consider that it is based on understanding (Elby, 2001). Therefore, one of the keys to improving our students' learning is to promote appropriate epistemological attitudes. For example, helping students understand the importance of the consistency and coherence of knowledge, as well as the difference between rote learning and deeper understanding, is certainly a teaching objective in itself, as part of the development of quality science education. As teachers, recognizing the images of science that we transmit, the relationship of science with technology and society, how to evolve towards notions that foster learning among our students, their ways of working and gender perspectives, are fundamental aspects that we must take into account and explicitly address in our classes.

In this paper we report the results obtained from the implementation of the CLASS test among an significant group of high school physics teachers and future teachers in Uruguay. The overall test scores show that the teachers' scores are higher and with less dispersion between categories than those of future teachers who have a lower mean value as well as higher dispersion. Another important finding is that the categories that have the highest agreement with the experts' opinion are the same for both groups, particularly the categories *Sense making/Effort*, *Connection with the real world* and *Personal interest*. On the other hand, we found the greatest differences in the categories of *Applied conceptual understanding* and *Sophistication in problem solving*. These observations suggest that the latter are acquired during professional training and practice. At the opposite end is the category *Confidence in problem solving*, for which we found no significant difference between the responses of teachers and teacher-training students (71% and 67%, respectively). These percentages are among the lowest (in the case of teachers it is indeed the lowest of all categories) for both groups, in relation to the respective average.

We also observed that the group of teacher-training students and the group of students who pursue careers with emphasis in Physics and Mathematics present similar scores. Although the results of these groups of students are generally good, they are clearly below those presented by professors and researchers. Thus, it seems clear that in secondary education there is a need to focus on working on issues related to the epistemological attitudes and beliefs of our students.

Finally, our work leaves several open questions to be addresses in future research. At what stage do teacher-training students transform their epistemological beliefs to approach that of experts? How do these beliefs impact teacher-training? How can teacher-training and high school courses be rethought to positively impact the epistemological attitudes of students? Are there differences between the beliefs of physics teacher-training students and those of other specialization areas? If we are to have a higher-quality science education, these and other relevant questions must be adressed.

**Acknowledgements**

We thank ANII and CFE for the financial support to the project "Knowing and influencing the epistemological conceptions of future physics teachers" (FSED_3_2019_1_157320).

**References**

Adams, W. K., Perkins, K. K., Podolefsky, N. S., Dubson, M., Finkelstein, N. D., & Wieman, C. E. (2006). New instrument for measuring student beliefs about physics and learning physics: The Colorado Learning Attitudes about Science Survey. *Physical review special topics-physics education research*, *2*(1), 010101.

Balta, N., Cessna, S. G., & Kaliyeva, A. (2020). Surveying Kazakh high school students' attitudes and beliefs about physics and learning with the Colorado learning attitudes about science survey. *Physics Education*, *55*(6), 065019.

Bates, S. P., Galloway, R. K., Loptson, C., & Slaughter, K. A. (2011). How attitudes and beliefs about physics change from high school to faculty. *Physical Review Special Topics-Physics Education Research*, *7*(2), 020114.

Brewe, E., Traxler, A., De La Garza, J., & Kramer, L. H. (2013). Extending positive CLASS results across multiple instructors and multiple classes of Modeling Instruction. *Physical Review Special Topics-Physics Education Research*, *9*(2), 020116.

Cahill, M. J., McDaniel, M. A., Frey, R. F., Hynes, K. M., Repice, M., Zhao, J., & Trousil, R. (2018). Understanding the relationship between student attitudes and student learning. *Physical Review Physics Education Research*, *14*(1), 010107.

*Chabay, R. & Sherwood, B. (2006). Brief electricity and magnetism assessment. Physical Review Special Topics-Physics Education Research, 2(1), 7-13.*

Ding, L. (2013, January). A comparative study of middle school and high school students' views about physics and learning physics. In *AIP Conference Proceedings* (Vol. 1513, No. 1, pp. 118-121). American Institute of Physics.

Docktor, J. L., & Mestre, J. P. (2014). Synthesis of discipline-based education research in physics. *Physical Review Special Topics-Physics Education Research*, *10*(2), 020119.

Elby, A. (2001). Helping physics students learn how to learn. *American Journal of Physics*, *69*(S1), S54-S64.

Hake, R. R. (1998). Interactive-engagement versus traditional methods: A six-thousand-student survey of mechanics test data for introductory physics courses. *American journal of Physics*, *66*(1), 64-74.

Halloun, I. A., & Hestenes, D. (1985). The initial knowledge state of college physics students. *American journal of Physics*, *53*(11), 1043-1055.

Heron, P. R., & Meltzer, D. E. (2005). The future of physics education research: Intellectual challenges and practical concerns.

Hestenes, D., Wells, M., & Swackhamer, G. (1992). Force concept inventory. *The physics teacher*, *30*(3), 141-158.

Kohl, P. B., & Vincent Kuo, H. (2012). Chronicling a successful secondary implementation of Studio Physics. *American Journal of Physics*, *80*(9), 832-839.

Kontro, I., & Buschhüter, D. (2020). Validity of Colorado Learning Attitudes about Science Survey for a high-achieving, Finnish population. *Physical Review Physics Education Research, 16(2), 020104.*

Madsen, A., McKagan, S. B., & Sayre, E. C. (2015). How physics instruction impacts students' beliefs about learning physics: A meta-analysis of 24 studies. *Physical Review Special Topics-Physics Education Research*, *11*(1), 010115.

Maloney, D. P., O'Kuma, T. L., Hieggelke, C. J., & Van Heuvelen, A. (2001). Surveying students' conceptual knowledge of electricity and magnetism. *American Journal of Physics, 69(S1), S12-S23.*

Milner-Bolotin, M., Antimirova, T., Noack, A., & Petrov, A. (2011). Attitudes about science and conceptual physics learning in university introductory physics courses. *Physical Review Special Topics-Physics Education Research*, *7*(2), 020107.

McDermott, L. C., & Redish, E. F. (1999). Resource letter: PER-1: Physics education research. *American journal of physics, 67(9), 755-767.*

Nissen, J. M., Horses, I. H. M., Van Dusen, B., Jariwala, M., & Close, E. W. (2021). Tools for identifying courses that support development of expertlike physics attitudes. *Physical Review Physics Education Research*, *17*(1), 013103.

Perkins, K. K., Adams, W. K., Pollock, S. J., Finkelstein, N. D., & Wieman, C. E. (2005, September). Correlating student beliefs with student learning using the Colorado Learning Attitudes about Science Survey. In *AIP Conference Proceedings* (Vol. 790, No. 1, pp. 61-64). American Institute of Physics.

Redish, E. F., Saul, J. M., & Steinberg, R. N. (1998). Student expectations in introductory physics. *American journal of physics*, *66*(3), 212-224.

Redish, E. F., & Hammer, D. (2009). Reinventing college physics for biologists: Explicating an epistemological curriculum. *American Journal of Physics*, *77*(7), 629-642.

Suwonjandee, N., Mahachok, T. & Asavapibhop, B. (2018). Evaluation of Thai students and teacher's attitudes in physics using Colorado Learning Attitudes about Science Survey (CLASS). *IOP Conf. Series: Journal of Physics: Conf. Series* 1144, 012124

Trowbridge, D. E., & McDermott, L. C. (1980). Investigation of student understanding of the concept of velocity in one dimension. *American journal of Physics*, *48*(12), 1020-1028.

Viennot, L. (1979). Spontaneous reasoning in elementary dynamics. *European journal of science education*, *1*(2), 205-221.

Wilcox, B. R., & Lewandowski, H. J. (2016). Students' epistemologies about experimental physics: Validating the Colorado Learning Attitudes about Science Survey for experimental physics. *Physical Review Physics Education Research*, *12*(1), 010123.

# ¿Cuáles son las actitudes y creencias en torno a la Ciencia de los profesores y estudiantes de profesorado de Física del Uruguay?

## Álvaro Suárez, Daniel Baccino, Martín Monteiro, Arturo C. Martí

Álvaro Suárez

ORCID https://orcid.org/0000-0002-5345-5565

Consejo de Formación en Educación, Instituto de Profesores Artigas, Montevideo, Uruguay

**alsua@outlook.com**

Daniel Baccino

ORCID https://orcid.org/0000-0001-5572-2623

Consejo de Formación en Educación, Instituto de Profesores Artigas, Montevideo, Uruguay

**dbaccisi@gmail.com**

Martín Monteiro

https://orcid.org/0000-0001-9472-2116

Universidad ORT Uruguay, Montevideo, Uruguay

**monteiro@ort.edu.uy**

Arturo C. Martí

https://orcid.org/0000-0003-2023-8676

Instituto de Física, Facultad de Ciencias, Udelar, Montevideo, Uruguay

**marti@fisica.edu.uy**

**Resumen:** Investigamos las concepciones epistemológicas de los profesores y futuros profesores de Física del Uruguay por medio de la aplicación del test CLASS (Colorado Learning Attitudes about Science Survey), uno de los instrumentos más aceptados en la comunidad de investigación en enseñanza de la Física. Los resultados obtenidos permiten comparar las actitudes y creencias en torno a la ciencias de ambos colectivos y evaluar en forma cuantitativa el acuerdo o no con las concepciones de expertos en la materia. Primero presentamos un pantallazo general de las respuestas y luego identificamos categorías en las que existen similitudes o diferencias significativas entre ambos colectivos estudiados y a su vez con la referencia de los expertos. Las categorías que muestran variaciones positivas o negativas significativas entre las opiniones de los profesores y futuros profesores indican las áreas donde la formación es favorable o desfavorable. Por otro lado, las áreas donde las diferencias con las opiniones de los expertos es globalmente notoria nos sugieren que deben ser fortalecidas en todos los ámbitos. Para tener una perspectiva más global, también comparamos nuestros resultados con algunos provistos por la literatura. Finalmente, dejamos una serie de preguntas que pensamos pueden favorecer indagaciones posteriores.

1. **Introducción.**

El famoso sicólogo estadounidense B. F. Skinner señaló en 1964 que "*la educación es aquello que perdura una vez que se olvidó todo lo aprendido*." Esta frase resume en forma ilustrativa que los contenidos específicos que enseñamos son solo una pequeña parte de la transformación que tiene lugar a medida que los estudiantes transitan por el sistema educativo. En efecto, los estudiantes no son "*cajas vacías*" donde los docentes depositan conocimientos, sino que vienen con toda una carga de conocimientos de la propia materia, una historia personal, una forma de interaccionar con sus compañeros y docentes, un posicionamiento frente al aprendizaje y un conjunto de expectativas de diversa índole entre otros aspectos. A menudo subestimados en relación a los disciplinares, dichos aspectos juegan un papel importante en la transformación que esperamos experimenten los estudiantes.

El conjunto de ideas, suposiciones y concepciones previas sobre la Ciencia, en especial su evolución, sus métodos, su validación o refutación se engloban en el concepto de *creencias epistemológicas*. Estas creencias, que en general no son explícitas, impactan fuertemente en la forma de enseñar y aprender (Redish, Saul y Steinberg, 1998). Por mencionar algunos aspectos, ¿consideramos la ciencia un conjunto de conocimientos grabados en piedra o un sistema en continua evolución? El rol de la experimentación, ¿es simplemente un requisito más para verificar las teorías o forma parte sustancial a la hora de entender la naturaleza, falsear teorías o delimitar su rango de validez? El posicionamiento frente a estas preguntas no es neutro. Si consideramos los conocimientos previos como inamovibles, no hace falta fomentar una actitud crítica ni interesarse por los nuevos avances. Asimismo, si consideramos que "todo vale" o que la evidencia experimental no es relevante caemos en el campo de las creencias seudocientíficas, o posturas tales como las sostenidas por terraplanistas o antivacunas.

Los docentes, en cada acción u omisión que realizamos en clase, estamos directa o indirectamente incidiendo sobre las actitudes y creencias de nuestros estudiantes en el aprendizaje y la enseñanza de la Física. Pensemos por ejemplo una situación hipotética, de una clase sobre cinemática donde resolvemos un problema en el pizarrón. Imaginemos que, como parte del procedimiento de resolución, escribimos las ecuaciones que describen un MRUV y les decimos a los estudiantes que para hallar la respuesta deben "*buscar la ecuación que tenga todos los datos excepto el pedido y simplemente despejar*". En una situación hipotética como la descrita, estaríamos sin quererlo, dándole una imagen errónea al estudiante de la manera en la se deben resolver los problemas en Física. En ese sentido, estamos formando (sin quererlo) una imagen incorrecta de la ciencia y de cómo aprender Física.

Los aspectos que enmarcan el conocimiento científico propio de la disciplina en la sociedad merecen ser discutidos, analizados y explicitados en las clases de ciencias. Es cierto que si bien la Filosofía como asignatura está presente en la mayoría de planes de estudio, muchos aspectos específicos de su relación con la Física deben ser discutidos junto con sus contenidos específicos. ¿Es posible hablar de inducción electromagnética sin mencionar el papel que juegan los motores y generadores en nuestra vida cotidiana? ¿Podemos presentar la mecánica cuántica como una teoría abstracta mientras sostenemos en la mano un teléfono inteligente con millones de transistores que se rigen justamente por las leyes de la mecánica cuántica? ¿Es razonable acaso estudiar los detalles de la fisión del uranio sin mencionar las implicaciones sociales y éticas de la bomba atómica? Estos ejemplos muestran la estrecha vinculación entre los contenidos que estudiamos y el conjunto de creencias y posturas en torno a la ciencia.

Existen otros aspectos más sutiles, no tan visibles, en torno a las actitudes frente a la ciencia que se discuten aún menos en nuestras clases. Algunos de estos involucran a las actitudes frente minorías o géneros y en especial al papel de las mujeres en relación a las "ciencias duras". ¿Por qué es menor el porcentaje de mujeres que eligen estas disciplinas? ¿La enseñanza que brindamos es "neutra" frente a estos aspectos? ¿En qué momento de la formación se origina esta aparente falta de interés? ¿Ocurre en los primeros años? ¿O en etapas más avanzadas? ¿Se debe a las condiciones labores? ¿Influye la imagen típica del científico que se presenta en la sociedad? ¿Se trata de una discriminación del estilo del "club de Tobi"[3]? Todas estas preguntas están cobrando mayor importancia hoy día. Para avanzar en la dirección de responder en forma convincente resulta natural indagar en las actitudes y creencias epistemológicas de los estudiantes y de los docentes.

Las actitudes y creencias tienen múltiples dimensiones. Una de ellas es la transversal que recorre los diferentes colectivos desde las posturas de la sociedad, de los estudiantes de ciencias, de los futuros docentes de Física, de los actuales docentes hasta los investigadores de la disciplina. En este trabajo nos enfocamos en evaluar y discutir las actitudes y creencias de los docentes de Física de enseñanza media y de los estudiantes de profesorado de Física en Uruguay. A continuación, en la próxima sección exponemos un conjunto de herramientas desarrolladas en los últimos años para indagar estos aspectos que constituyen nuestro marco teórico y presentamos los resultados más relevantes que se han obtenido de su aplicación en diversos contextos. Avanzando en los aspectos centrales de este trabajo, concernientes a las actitudes y creencias de los docentes uruguayos, en la sección 3, presentamos la metodología empleada y en la sección 4 los principales resultados. Finalmente, la discusión y consideraciones finales se expone en la sección 5.

## 2. Herramientas de análisis de las creencias epistemológicas y sus principales resultados

### 2.1 Cuestionarios estandarizados en la investigación en Enseñanza de la Física

En diferentes ámbitos de la Física y de su enseñanza se comenzó a observar desde hace casi medio siglo que frecuentemente los estudiantes aprendían a resolver los problemas y aprobaban sus cursos, pero, al enfrentarse a problemas conceptuales sencillos, mostraban no tener una comprensión cabal de los fenómenos estudiados (Viennot, 1979; Trowbridge y McDermott, 1980; Halloun y Hestenes, 1985; Hake, 1996, Docktor y Mestre, 2014). A partir de estas observaciones, surgieron voces desde el núcleo de la Física que reclamaban la necesidad de cambios. A lo largo de los años siguientes los estudios en esta temática fueron adquiriendo coherencia surgiendo un nuevo campo con identidad propia: la investigación en enseñanza de la Física (PER por sus siglas en inglés) (McDermott, 1999) que apunta a generar conocimiento basado en métodos científicos ampliamente aceptados y en especial en contar con resultados cuantitativos con respaldo estadístico sólido (Docktor y Mestre, 2014).

Uno de los pilares de la PER es la evaluación sistemática de conocimientos. En este marco se idearon diversos cuestionarios para la evaluación de conocimientos destacándose el *Force Concept Inventory* (Hestenes, Wells y Swackhamer, 1992). Este cuestionario, aplicado a miles de estudiantes en cientos de universidades, apunta a evaluar el pasaje entre una concepción aristotélica de la mecánica a una newtoniana. La principal conclusión derivada del análisis de los resultados es que los métodos de enseñanza activa, en los que los estudiantes se involucran activamente, redundan en mejores aprendizajes en comparación con los métodos tradicionales, principalmente clases magistrales, donde el estudiante tiene una actitud pasiva. Esta mayor "ganancia" en el aprendizaje, es un resultado robusto en términos estadísticos, no depende del estudiante, del docente ni del contexto donde ocurre el aprendizaje (Hake, 1998).

En los años posteriores se propusieron numerosos cuestionarios, como el *Conceptual Survey of Electricity and Magnetism* (CSEM), Malonney et al. (2001) o el *Brief electricity and magnetism assesssment* (BEMA), Chabay y Sherwood (2006), que apuntan a otros campos de la Física. Si bien en los comienzos la elaboración de estos cuestionarios era relativamente artesanal, con el pasar del tiempo se definieron una serie de pasos, requisitos estadísticos y etapas que debe cumplir cada propuesta para ser aceptada en la comunidad (Heron y Meltzer, 2005). Paralelamente, en varios sitios colaborativos (el más reconocido es http://www.physport.org) se publican los cuestionarios, generalmente de libre acceso, con enlaces a los artículos, información complementaria como traducciones a muchos idiomas, planillas de cálculo y otras herramientas para facilitar su implementación. Por otro, cuando se cumplen estos requisitos es posible compartir la información, tanto subir nuestros resultados como acceder a los de otras instituciones.

---

[3] Expresión tomada de la popular historieta "La pequeña Lulu" y utilizada en América Latina para aludir a situaciones o contextos donde quiénes asisten y participan son exclusivamente personajes varones.

**2.2 El análisis de las actitudes y creencias epistemológicas.**

El campo de las actitudes y creencias epistemológicas no escapa a la pretensión de los investigadores de obtener información cuantitativa. En ese sentido, se han desarrollado en los últimos veinticinco años una serie de cuestionarios o pruebas estandarizadas para evaluar las actitudes y creencias epistemológicas de los estudiantes sobre la Física, su enseñanza y aprendizaje. A diferencia de los test de conocimientos específicos, basados mayoritariamente en preguntas múltiple opción, los test de actitudes y creencias suelen proponer a los estudiantes que indiquen su grado de acuerdo o desacuerdo (conocido como escala de Likert) con afirmaciones que reflejan (o no) la opinión de "expertos" de la disciplina (es decir, físicos profesionales, investigadores). Cada afirmación tiene un valor esperado que es asignado durante el proceso de diseño y validación del instrumento, en base a repetidas interacciones con expertos. Las diferencias entre las respuestas de los estudiantes y las respuestas de los expertos, constituye la materia prima para analizar el estado epistemológico del grupo de estudiantes. En general es de interés analizar el impacto que determinados cursos y enfoques de enseñanza tienen en las actitudes y creencias de los estudiantes. Con este fin se aplica la prueba al menos dos veces, la primera al comenzar el curso (pretest) y la segunda al terminarlo (postest). La diferencia entre ambas pruebas permite medir los cambios en las actitudes y creencias en función de elementos demográficos, metodologías y estrategias de enseñanza.

Dentro del conjunto de pruebas estandarizadas se destacan especialmente tres: el MPEX (*Maryland Physics Expectations Survey*) (Redish, et al., 1998), que como su nombre lo indica, apunta a sondear las expectativas de los estudiantes en relación a la Física, el CLASS (*Colorado Learning Attitudes about Science Survey*) (Adams, et al., 2006), que apunta a evaluar las creencias de los estudiantes sobre la Física y sobre su aprendizaje y el E-CLASS (*Colorado Learning Attitudes about Science Survey for Experimental Physics*) (Wilcox y Lewandowski, 2016) que sigue la misma orientación del CLASS pero apunta a los aspectos experimentales de Física.

**2.3. Métodos de enseñanza y posturas epistemológicas**

En la literatura de PER se han publicado numerosos trabajos indagando en la relación entre los métodos y resultados de la enseñanza y las posturas epistemológicas de los estudiantes (Madsen, McKagan y Sayre, 2015). En particular se ha mostrado que ciertas posturas epistemológicas "negativas" de los estudiantes son sin duda un obstáculo para el desarrollo de una educación científica de calidad pues afectan la manera en que los alumnos aprenden y abordan los cursos de ciencias (Perkins, et al., 2005; Milner-Bolotin, Antimirova, Noack y Petrov, 2011). Por ejemplo, las maneras que un estudiante cree que debe aprender Física, sus prácticas metacognitivas, así como la imagen que tiene de la Física como ciencia, son algunos de los aspectos del currículum oculto que deberían afectar la forma de aprender. Un estudiante que cree que la Física consiste principalmente en hechos y fórmulas desconectadas, estudiará de manera diferente a otro que la ve como una red de conceptos interconectados. Aquellos que ven el conocimiento de la Física como una red coherente de ideas, tienen motivos para realizar prácticas metacognitivas para monitorear su aprendizaje (Redish, Saul y Steinberg, 1998). En relación con el rendimiento académico, diversos estudios han mostrado la existencia de una correlación positiva con las puntuaciones del pretest del CLASS y el MPEX (Perkins, et al., 2005; Cahill, et al., 2018). En los últimos años, se han realizado múltiples investigaciones cuantitativas sobre los cambios que se producen en las actitudes y creencias de los estudiantes en función de su formación previa, tipos de cursos y estrategias de enseñanza utilizadas, así como la relación entre el pretest y el rendimiento académico, entre otras variables (Madsen, McKagan y Sayre, 2015).

Del conjunto de investigaciones, queremos destacar particularmente aquellas que estudian los cambios en las posturas epistemológicas en función de los métodos de enseñanza, así como las que analizan la relación entre dichas posturas y el entendimiento conceptual. A partir de un meta-análisis realizado examinando 24 trabajos de investigación publicados, Madsen, McKagan y Sayre (2015) clasificaron las estrategias de enseñanza de los cursos de Física en tres categorías en función de los cambios en las posturas epistemológicas de los estudiantes:

A) Cursos donde se produce un deterioro de las actitudes y creencias de los estudiantes (el resultado promedio del postest sobre actitudes y creencias es menor al del pretest). En esta categoría se encuentran los cursos enfocados con metodologías de enseñanza tradicionales, así como aquellos desarrollados con algunas de las metodologías de enseñanza activa más difundidas, tales como la "Instrucción de pares" y "Tutoriales de Física Introductoria" (Adams et al, 2006; Madsen, et al., 2015). Aunque el hecho de que un curso tradicional no promueva las actitudes y creencias adecuadas podría ser algo previsible, a priori resulta llamativo que, para cursos basados en algunas estrategias de enseñanza activas, los estudiantes también tengan un desempeño menor en el CLASS y el MPEX después de haber transitado por el curso.

B) Cursos donde no se detecta una diferencia significativa en las actitudes y creencias de los estudiantes antes y después del curso. Aquí se encuentran los cursos donde se toman en cuenta algunos aspectos epistemológicos en su implementación (Kohl y Vincent Kuo, 2012; Madsen, et al., 2015). En estos cursos se implementan estrategias de enseñanza que promueven el desarrollo de habilidades de razonamiento, la realización de inferencias a partir de las observaciones y reflexiones sobre por qué confiamos en las ideas científicas, entre otras.

C) Cursos donde se detecta una mejora significativa en las actitudes y creencias de los estudiantes. En este grupo se encuentran dos tipos: por un lado, aquellos que son radicalmente reestructurados, basados en la construcción de modelos (Brewe, et al., 2013) y por otro, los que tienen un fuerte énfasis en aspectos epistemológicos (Elby, 2001; Redish y Hammer, 2009). En los cursos basados en la construcción de modelos, los estudiantes trabajan en pequeños grupos, realizan experimentos y analizan los resultados con el fin de elaborar modelos de los diferentes fenómenos. Un aspecto central de este tipo de estrategias es la puesta en común de las conclusiones de cada equipo y la promoción de discusiones con otros compañeros de forma de permitir a los estudiantes validar y refinar los modelos construidos. Por otro lado, en los cursos con un fuerte énfasis en lo epistemológico, se desarrollan actividades donde se promueve la metacognición, y se trabaja por ejemplo con tutoriales que enfatizan la reconciliación entre las ideas intuitivas y pensamiento científico formal.

Vemos entonces que, como se desprende de la literatura, exceptuando aquellos cursos pensados específicamente para promover mejoras en las actitudes y creencias de los estudiantes, en general los alumnos las deterioran como efecto colateral de los cursos de Física. Sorpresivamente, al finalizar muchos de los cursos de Física, más alumnos creen por ejemplo que la Física está menos conectada con el mundo real, es menos coherente, el razonamiento es menos importante y el aprendizaje memorístico resulta útil. Estos resultados, a priori sumamente chocantes, nos interpelan sobre la promoción en forma involuntaria de actitudes y creencias contrarias a las que esperaríamos se desarrollaran de manera natural por el solo tránsito a través de los cursos. La estrecha relación entre posturas epistemológicas y resultados del proceso educativo justifica ampliamente la necesidad de conocer y actuar sobre las actitudes y creencias de los estudiantes a lo largo de todo el sistema educativo.

## 3. Metodología

En la investigación que presentamos en este trabajo optamos por utilizar CLASS, la encuesta de actitudes de aprendizaje de la ciencia desarrollada por el grupo de PER de la Universidad de Colorado Boulder, en Estados Unidos (Adams, et al., 2006). Este cuestionario cuenta con muy amplia aceptación dado que ha pasado un exhaustivo proceso de validación. Para su creación se partió de herramientas como el MPEX y VASS para luego revisarlo en base a entrevistas con estudiantes y expertos (16 físicos con amplia experiencia en la enseñanza que estuvieron de acuerdo con las respuestas para casi todas las preguntas). Las categorías de CLASS se crearon utilizando un análisis factorial de base reducida, donde las categorías estadísticas en bruto y las categorías predeterminadas por los investigadores se combinaron de forma iterativa. Finalmente, fue aplicado a miles de estudiantes y se verificó que aquellos con más experiencia en física, tenían creencias más parecidas a las de los expertos. Los resultados de su aplicación en diferentes contextos geográficos y sociales o en grupos donde se aplican diferentes estrategias de enseñanza se encuentran disponibles para su análisis y comparación (Milner-Bolotin, Antimirova, Noack y Petrov, 2011; Ding, 2013; Suwonjandee, Mahachok y Asavapibhop, 2018; Balta, Cessna, Kaliyeva, 2020, Kontro y Buschhüter, 2020; Nissen et al., 2021). Este instrumento valora las actitudes de los estudiantes para aprender Física, cómo piensan que se relaciona la Física con su vida cotidiana y qué piensan sobre la Física. Está conformado por 42 afirmaciones que se presentan cada una con escala de Likert de cinco niveles, desde completamente en desacuerdo (1), hasta completamente de acuerdo (5). Anotamos algunas preguntas a modo de ejemplo:

- *"Yo pienso en la Física que se involucra en mi vida cotidiana."*
- *"El conocimiento en Física consiste de muchos temas desconectados."*
- *"Después de estudiar un tema de Física y siento que lo entiendo, tengo dificultades para resolver problemas sobre el mismo tema"*
- *"Cuando resuelvo un problema de Física, busco una ecuación que utiliza las variables dadas en el problema y substituyo los valores"*

Las afirmaciones se pueden organizar, según los autores de CLASS, en ocho categorías (de forma no exclusiva), tal como se muestra en la tabla 1. Al procesar una encuesta realizada por un estudiante, cada una de sus respuestas recibirá un valor (-1 o +1), según el grado de desacuerdo (-1) o acuerdo (+1) entre la respuesta dada por el estudiante y la respuesta típica de los expertos. Este criterio se aplica solamente a las 36 preguntas en las que existe real consenso entre los expertos (solo 27 de las 36 están categorizadas). Por el contrario, no se aplica a las seis preguntas (ver la última fila de la tabla 1) que no muestran acuerdo entre expertos. A partir de estas valoraciones se determinan los niveles de acuerdo ("favorable" y "desfavorable") para cada una de las ocho categorías definidas por los autores, así como para la encuesta en su totalidad. Este proceso se aplica a todos los individuos del grupo analizado y finalmente se calculan los promedios para cada afirmación, categoría y el total de la encuesta.

| Categoría | Afirmaciones | ID |
|---|---|---|
| Conexión con el mundo real (Real world conection) | 28, 30, 35, 37 | RWC |
| Interés personal (Personal interest) | 3, 11, 14, 25, 28, 30 | PI |
| Hacer sentido / esfuerzo (Sense making effort) | 11, 23, 24, 32, 36, 39, 42 | SM/E |
| Entendimiento conceptual (Conceptual comprehension) | 1, 5, 6, 13, 21, 32 | CC |
| Entendimiento conceptual aplicado (Applied conceptual undestanding) | 1, 5, 6, 8, 21, 22, 40 | ACU |
| Resolución de problemas. General (Problem solving, general) | 13, 15, 16, 25, 26, 34, 40. | PSG |
| Resolución de problemas. Confianza (Problem solving, confidence) | 15, 16, 34, 40 | PSC |
| Resolución de problemas. Sofisticación (Problem solving sophistication) | 5, 21, 22, 25, 34, 40 | PSS |
| Sin acuerdo entre expertos | 4, 7, 9, 31, 33, 41 | |

**Tabla 1. Identificación de categorías y las declaraciones comprendidas en cada una de ellas (Traducción de Adams et al, 2006, de los autores)**

La propuesta del test se implementó a través un formulario electrónico, enviado mediante convocatorias abiertas, a profesores de Física con actividad en educación media, y estudiantes de primer año del profesorado de Física del CFE (Consejo de Formación en Educación de Uruguay). La encuesta estuvo abierta para cada colectivo, docentes y estudiantes de todo Uruguay, durante un mes en la primera mitad del año 2020.

Recibimos 143 respuestas de profesores. El género declarado por los encuestados fue: Mujer (46% del total), Hombre (52%), no lo declaró un 2%. El 52% de los encuestados indica que desarrolla su actividad docente en Montevideo (la capital del país), mientras que el 48% restante se distribuye en otros departamentos del país. En cuanto a la edad de los encuestados, 31% declara entre 31 y 40 años, 22% se declaran en las franjas etarias contiguas a la anterior (18-30 y 41-50), 25% registra más de 51 años de edad.[4]

Un total de 138 estudiantes de casi todos los centros en los que se imparte el profesorado de Física contestaron la encuesta. Con respecto al género, 62% de los encuestados declaran ser mujer, 38% declara ser hombre, y ningún encuestado tomó las otras opciones disponibles. En términos del Instituto donde están radicados, la mayoría de ellos (56%) cursa en la modalidad semipresencial[5], mientras que el resto lo hace en forma presencial en diferentes centros del país.

Los resultados de cada una de las encuestas se analizaron en base a las categorías definidas y validadas por los autores del instrumento. Para cuantificar los resultados, con ese punto de partida, utilizamos una planilla electrónica desarrollada por los autores (extraída del portal de Physport) y adaptada por los autores de este trabajo. En la sección siguiente mostramos los principales resultados obtenidos.

## 4. Análisis de los resultados

4.1 Profesores y estudiantes de profesorado del Uruguay

En un primer análisis de los resultados, damos una mirada a todas las respuestas categorizadas de los profesores. En el histograma de la figura 1 se representan las frecuencias (cantidad de profesores) cuyas respuestas se encuentran de acuerdo con los expertos, en intervalos de ancho 10 puntos porcentuales. Por ejemplo, un total de 56 profesores dieron respuestas "favorables" que están entre 80% y 90%. La forma típica de caracterizar a esa distribución es mediante el promedio, que en este caso es 80%, la desviación estándar (13%) y la desviación estándar del promedio (1%).

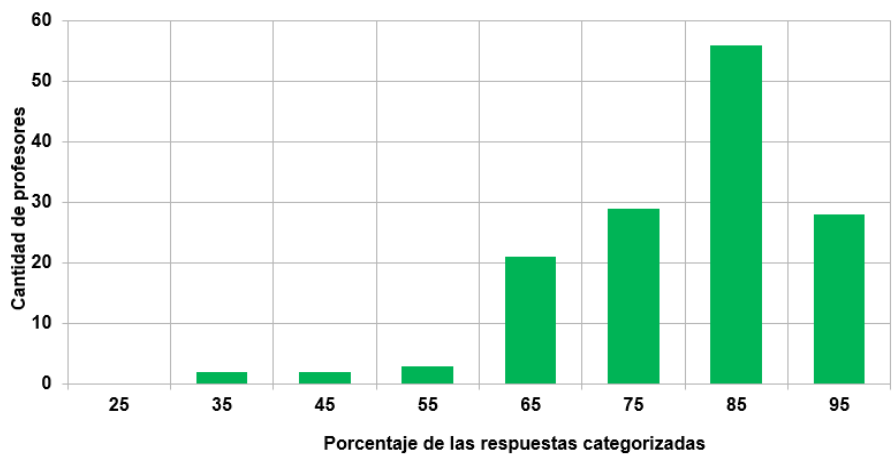

Figura 1. Resultados globales de las respuestas a las preguntas categorizadas en el test CLASS por parte de 143 profesores de Física de Uruguay.

En la figura 2 mostramos los resultados por categorías en la encuesta a profesores, para respuestas favorables, neutras y no favorables (según el ID de la tabla 1); las dos primeras columnas representan los porcentajes correspondientes a todas las respuestas (All) y al total de respuestas categorizadas (All C.) respectivamente.

---

[4] Se puede acceder a los resultados de la encuesta a profesores y estudiantes a través de los siguientes enlaces: bit.ly/CLASS-profes-UY2020 y bit.ly/CLASS-formacion-UY2020

[5] Semipresencial consiste en una *modalidad de cursado* de la carrera de profesorado, habilitada (desde el año 2003) para los Centros de Formación Docente de todo el Uruguay, que no cuenten en su oferta con las asignaturas específicas en algunas Especialidades, entre las que cuenta el profesorado de Física.

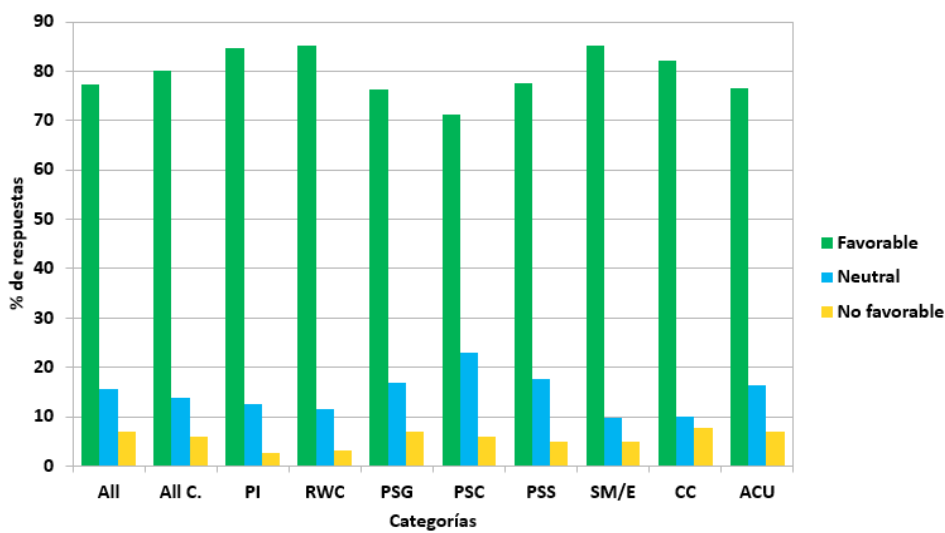

Figura 2. Desempeño de los profesores de Física encuestados con el test CLASS desglosado por categorías.

El gráfico de la figura 3 muestra los resultados de respuestas favorables de los profesores y de los estudiantes de profesorado de Física discriminado por categorías de CLASS (corresponden idénticos comentarios sobre la identificación de categorías que las realizadas sobre el gráfico de la figura 2). Se representa, para cada categoría y en ambos grupos, el error estándar del promedio. El eje de porcentajes se define entre valores de 45% y 95%, con el objetivo de visualizar mejor la zona de interés.

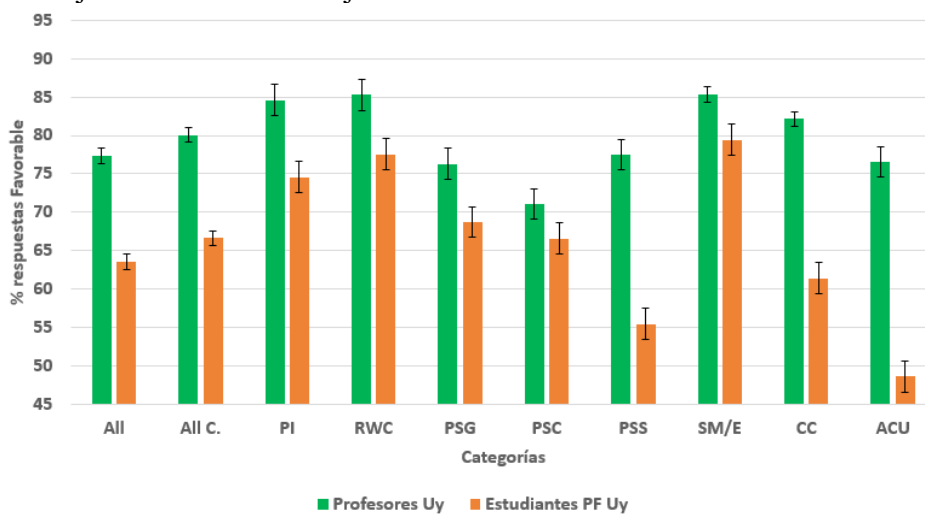

Figura 3. Desempeño de 138 estudiantes de profesorado de Física (Estudiantes PF Uy) y 143 profesores de Física en actividad (Profesores Uy) en el test CLASS desglosado por categoría.

Un aspecto que resalta a la vista en el gráfico es la diferencia en los promedios de las respuestas de profesores y estudiantes. En el caso de profesores el promedio de respuestas categorizadas es 80%, mientras que en estudiantes es 67%. Si consideramos el rango definido por los promedios de todas las categorías, observamos que en los estudiantes es de aproximadamente 30 puntos porcentuales, mientras que en caso de los profesores es aproximadamente 15 puntos. En el caso de los profesores las respuestas a las categorías están menos dispersas que en los subconjuntos categorizados para los estudiantes.

Las categorías en las que hay más acuerdo con los expertos son compartidas por estudiantes y profesores. En algunas categorías el acuerdo de estudiantes y profesores con expertos es muy similar, por ejemplo, en

*Esfuerzo/Hacer sentido* y *Confianza en la Resolución de problemas*. En el otro extremo, donde se visualizan diferencias mayores, se encuentran las categorías correspondientes al *Entendimiento conceptual aplicado* y *Sofisticación en la resolución de problemas*. Asumiendo que es posible considerar a los dos grupos de individuos encuestados como parte de una mirada pseudo-longitudinal, una hipótesis que puede explicar estas diferencias, refiere a los procesos de formación sistemática y los mediados por la práctica cotidiana de los docentes.

Propusimos la encuesta a 79 estudiantes que ingresaron a las Licenciaturas en Física y Matemática (Facultad de Ciencias, Udelar) y a 105 estudiantes que ingresaron a la Facultad de Ingeniería (Udelar). No encontramos diferencias significativas entre las respuestas de los estudiantes universitarios y los estudiantes de profesorado de Física encuestados. La misma encuesta se propuso a investigadores locales en el área de las Ciencias Físicas. Al comparar estos últimos resultados con los de los profesores de física, no encontramos diferencias significativas. Los resultados del análisis de las respuestas de estos colectivos serán analizados en otro trabajo.

### 4.2 Comparación con resultados internacionales

Es interesante comparar los resultados globales de los estudiantes de profesorado de Física de Uruguay, con estudiantes universitarios de otras partes del mundo. Utilizando datos del meta-análisis realizado por Madsen, McKagan y Sayre (2015) y de Bates, Galloway, Loptson y Slaughter (2011), tomamos los valores promedio de la implementación del test CLASS a estudiantes de trece universidades de Canadá, Estados Unidos y el Reino Unido, con la particularidad que el perfil de las carreras de dichos estudiantes es tal que toman cursos de Física basados en cálculo. Este punto es de particular importancia, ya que según la literatura (Madsen, McKagan y Sayre, 2015), los estudiantes que realizan carreras con fuerte énfasis en Física y Matemática, obtienen mejores resultados en el CLASS.

Mientras que el promedio del test CLASS a los estudiantes de formación docente para las 36 preguntas donde hay acuerdo entre expertos fue de 63,5%, el promedio para los trece grupos es de 64,1% con una desviación estándar de 6,0%. Estos resultados, muestran cómo los estudiantes del profesorado de Física de Uruguay, obtienen resultados similares a las poblaciones utilizadas para la comparación. Aunque no nos dicen nada respecto a los sistemas educativos, sugieren que quizás haya una tendencia en los resultados promedio del test CLASS para estudiantes con perfiles hacia el área de ciencias e ingeniería, independientemente de su país de origen.

Por otra parte, en relación a las actitudes y creencias de los profesores de física, puede resultar constructiva la comparación con un estudio realizado en Tailandia, un país con características demográficas, sociales, económicas y culturales muy diferentes a las de Uruguay. En el estudio realizado por Suwonjandee, Mahachok y Asavapibhop (2018), se analizaron las respuestas de 196 profesores de física y 211 estudiantes de secundaria. A pesar de las diferencias mencionadas entre los contextos de ambos colectivos docentes, las respuestas fueron muy similares en la mayoría de las preguntas y categorías del CLASS. Solamente en dos de las 42 preguntas del test se evidencian diferencias significativas entre los profesores de Tailandia y de Uruguay (preguntas 8 y 22). Estas son dos de las seis preguntas de la categoría *Entendimiento conceptual aplicado*.

En las preguntas mencionadas, los profesores de Uruguay mostraron un mayor grado de acuerdo con los expertos, que los profesores de Tailandia. Algo muy similar a lo que ocurre con los estudiantes de profesorado de física de Uruguay, que también muestran grandes diferencias con los expertos precisamente en estas mismas dos preguntas. De hecho, puede ser relevante señalar que estas preguntas son las que en nuestra investigación presentan el mayor grado de discrepancia entre las respuestas de los profesores de física y los estudiantes de profesorado de física:

- Pregunta 8: "*Cuando resuelvo un problema de física, busco una ecuación que utiliza las variables dadas en el problema y sustituyo los valores.*" El grado de acuerdo con los expertos fue de 60%, 24% y 9%, para profesores de Uruguay, estudiantes de profesorado de física de Uruguay y profesores de Tailandia, respectivamente.

- Pregunta 22: "*Si quiero aplicar un método usado en un problema de física en otro problema, estos problemas deben involucrar situaciones muy similares.*" El grado de acuerdo con los expertos fue de 67%, 27% y 38%, para profesores de Uruguay, estudiantes de profesorado de física de Uruguay y profesores de Tailandia, respectivamente.

Estas dos preguntas parecen señalar que la actitud frente a los problemas de física es un aspecto clave en la formación de los profesores de física y parecen ser una fortaleza relativa de los profesores uruguayos. El bajo grado de acuerdo que los estudiantes de profesorado de física de Uruguay muestran en la pregunta 8, parece sugerir que abordan los problemas de física conectando ecuaciones, resolviéndolas con los datos disponibles, sin pensar demasiado en los aspectos conceptuales. La pregunta 22 parece reforzar esta idea, que los estudiantes de profesorado no logran comprender plenamente los aspectos conceptuales y los principios básicos de la física, y en su lugar aprenden física memorizando ecuaciones, lo cual les impide abordar nuevas situaciones con la confianza necesaria para resolverlas.

## 5. Comentarios finales

Las actitudes y creencias epistemológicas de los profesores y estudiantes juegan un rol fundamental en el aula, involucrando aspectos muy disímiles. En ese sentido, ser conscientes de su importancia y reconocer el impacto que pueden tener nuestras acciones en el aula en la imagen de ciencia que promovemos, la elección de carreras futuras o los logros académicos de nuestros estudiantes, resulta imprescindible.

Las creencias "epistemológicas" de los estudiantes, sus puntos de vista sobre la naturaleza del conocimiento y el aprendizaje, son un obstáculo para el desarrollo de una educación científica de calidad, afectando la manera en que aprenden y abordan los cursos de ciencias. Por ejemplo, quienes consideran que el aprendizaje consiste principalmente en absorber información, tendrán una postura diferente frente al conocimiento que aquellos que consideran que se basa en la comprensión (Elby, 2001). Por ello, una de las claves para mejorar los aprendizajes de nuestros estudiantes es promover posturas epistemológicas adecuadas. Ayudar a los estudiantes, por ejemplo, a comprender la importancia de la consistencia y coherencia del conocimiento, así como la diferencia entre la memorización y una comprensión más profunda, es sin duda un objetivo de enseñanza en sí mismo, como parte del desarrollo de una educación científica de calidad. Reconocer como docentes, las imágenes de ciencia que transmitimos, su relación con la tecnología y la sociedad, cómo evolucionar hacia concepciones que favorezcan el aprendizaje de nuestros estudiantes, sus formas de trabajar y las perspectivas de género, son aspectos fundamentales que debemos tener en cuenta y abordar explícitamente en nuestras clases.

En este trabajo reportamos los resultados obtenidos de la implementación de CLASS a un conjunto importante de profesores y futuros profesores de Física de enseñanza media de Uruguay. El puntaje global de los test muestra que el resultado de los profesores es más alto y con menos dispersión entre categorías que los de los futuros profesores que tienen un valor medio menor y además una mayor dispersión. Otro aspecto importante es que en ambos conjuntos las categorías que tienen mayor acuerdo con la opinión de los expertos son coincidentes, en particular en las de *Esfuerzo/Hacer sentido, Conexión con el mundo real e Interés personal*. Por otro lado, en las categorías de *Entendimiento conceptual aplicado* y *Sofisticación en la resolución de problemas* constatamos las mayores diferencias. Estas observaciones sugieren que estos últimos aspectos son adquiridos durante la formación y práctica profesional. En el extremo opuesto se encuentra la categoría *Confianza en la resolución de problemas*, donde no hemos encontrado diferencia significativa entre las respuestas de los grupos de profesores (71 %) y estudiantes de profesorado (67 %). Estos porcentajes se encuentran entre los más bajos (en el caso de los profesores es el más bajo de todas las categorías) para ambos grupos, en relación al promedio respectivo.

Observamos también que la comparación del grupo de estudiantes de profesorado con el grupo de estudiantes que optan por carreras con énfasis en Física y Matemática, presentan puntajes similares. Aunque los resultados de estos grupos de estudiantes son en general buenos, se encuentran claramente por debajo de los presentados por profesores e investigadores. De esta manera, parece clara la necesidad de apostar en enseñanza media a trabajar sobre cuestiones vinculadas a las actitudes y creencias epistemológicas de nuestros estudiantes.

Finalmente, el trabajo realizado deja varias preguntas abiertas por responder en futuras investigaciones. ¿En qué etapa los estudiantes de profesorado transforman sus creencias epistemológicas, acercándose a la de expertos? ¿De qué manera estas creencias impactan en la formación docente? ¿Cómo se pueden repensar los cursos de profesorado y de enseñanza media para impactar de manera positiva en las posturas epistemológicas de los estudiantes? ¿Existen diferencias entre las creencias de los estudiantes del profesorado de Física con las de otras especialidades? Si queremos tener una educación científica de mayor calidad, éstas y otras preguntas relevantes, deben ser investigadas.

## 7. Referencias

Adams, W. K., Perkins, K. K., Podolefsky, N. S., Dubson, M., Finkelstein, N. D., & Wieman, C. E. (2006). New instrument for measuring student beliefs about physics and learning physics: The Colorado Learning Attitudes about Science Survey. *Physical review special topics-physics education research*, *2*(1), 010101.

Balta, N., Cessna, S. G., & Kaliyeva, A. (2020). Surveying Kazakh high school students' attitudes and beliefs about physics and learning with the Colorado learning attitudes about science survey. *Physics Education*, *55*(6), 065019.

Bates, S. P., Galloway, R. K., Loptson, C., & Slaughter, K. A. (2011). How attitudes and beliefs about physics change from high school to faculty. *Physical Review Special Topics-Physics Education Research*, *7*(2), 020114.

Brewe, E., Traxler, A., De La Garza, J., & Kramer, L. H. (2013). Extending positive CLASS results across multiple instructors and multiple classes of Modeling Instruction. *Physical Review Special Topics-Physics Education Research*, *9*(2), 020116.

Cahill, M. J., McDaniel, M. A., Frey, R. F., Hynes, K. M., Repice, M., Zhao, J., & Trousil, R. (2018). Understanding the relationship between student attitudes and student learning. *Physical Review Physics Education Research*, *14*(1), 010107.

*Chabay, R. & Sherwood, B. (2006). Brief electricity and magnetism assessment. Physical Review Special Topics-Physics Education Research, 2(1), 7-13.*

Ding, L. (2013, January). A comparative study of middle school and high school students' views about physics and learning physics. In *AIP Conference Proceedings* (Vol. 1513, No. 1, pp. 118-121). American Institute of Physics.

Docktor, J. L., & Mestre, J. P. (2014). Synthesis of discipline-based education research in physics. *Physical Review Special Topics-Physics Education Research*, *10*(2), 020119.

Elby, A. (2001). Helping physics students learn how to learn. *American Journal of Physics*, *69*(S1), S54-S64.

Hake, R. R. (1998). Interactive-engagement versus traditional methods: A six-thousand-student survey of mechanics test data for introductory physics courses. *American journal of Physics*, *66*(1), 64-74.

Halloun, I. A., & Hestenes, D. (1985). The initial knowledge state of college physics students. *American journal of Physics*, *53*(11), 1043-1055.

Heron, P. R., & Meltzer, D. E. (2005). The future of physics education research: Intellectual challenges and practical concerns.

Hestenes, D., Wells, M., & Swackhamer, G. (1992). Force concept inventory. *The physics teacher*, *30*(3), 141-158.

Kohl, P. B., & Vincent Kuo, H. (2012). Chronicling a successful secondary implementation of Studio Physics. *American Journal of Physics*, *80*(9), 832-839.

Kontro, I., & Buschhüter, D. (2020). Validity of Colorado Learning Attitudes about Science Survey for a high-achieving, Finnish population. *Physical Review Physics Education Research, 16(2), 020104.*

Madsen, A., McKagan, S. B., & Sayre, E. C. (2015). How physics instruction impacts students' beliefs about learning physics: A meta-analysis of 24 studies. *Physical Review Special Topics-Physics Education Research*, *11*(1), 010115.

Maloney, D. P., O'Kuma, T. L., Hieggelke, C. J., & Van Heuvelen, A. (2001). Surveying students' conceptual knowledge of electricity and magnetism. *American Journal of Physics, 69(S1), S12-S23.*

Milner-Bolotin, M., Antimirova, T., Noack, A., & Petrov, A. (2011). Attitudes about science and conceptual physics learning in university introductory physics courses. *Physical Review Special Topics-Physics Education Research*, *7*(2), 020107.

McDermott, L. C., & Redish, E. F. (1999). Resource letter: PER-1: Physics education research. *American journal of physics, 67(9), 755-767.*

Nissen, J. M., Horses, I. H. M., Van Dusen, B., Jariwala, M., & Close, E. W. (2021). Tools for identifying courses that support development of expertlike physics attitudes. *Physical Review Physics Education Research*, *17*(1), 013103.

Perkins, K. K., Adams, W. K., Pollock, S. J., Finkelstein, N. D., & Wieman, C. E. (2005, September). Correlating student beliefs with student learning using the Colorado Learning Attitudes about Science Survey. In *AIP Conference Proceedings* (Vol. 790, No. 1, pp. 61-64). American Institute of Physics.

Redish, E. F., Saul, J. M., & Steinberg, R. N. (1998). Student expectations in introductory physics. *American journal of physics*, *66*(3), 212-224.

Redish, E. F., & Hammer, D. (2009). Reinventing college physics for biologists: Explicating an epistemological curriculum. *American Journal of Physics*, *77*(7), 629-642.

Suwonjandee, N., Mahachok, T. & Asavapibhop, B. (2018). Evaluation of Thai students and teacher's attitudes in physics using Colorado Learning Attitudes about Science Survey (CLASS). *IOP Conf. Series: Journal of Physics: Conf. Series* 1144, 012124

Trowbridge, D. E., & McDermott, L. C. (1980). Investigation of student understanding of the concept of velocity in one dimension. *American journal of Physics*, *48*(12), 1020-1028.

Viennot, L. (1979). Spontaneous reasoning in elementary dynamics. *European journal of science education*, *1*(2), 205-221.

Wilcox, B. R., & Lewandowski, H. J. (2016). Students' epistemologies about experimental physics: Validating the Colorado Learning Attitudes about Science Survey for experimental physics. *Physical Review Physics Education Research*, *12*(1), 010123.